# Superionic surface Li-ion transport in carbonaceous materials

Jianbin Zhou,[1#] Shen Wang,[1#] Chaoshan Wu,[2#] Ji Qi,[1#] Hongli Wan,[3#] Shen Lai,[4] Shijie Feng,[1] Tsz Wai Ko,[1] Zhaohui Liang,[1] Ke Zhou,[1] Nimrod Harpak,[1] Nick Solan,[1] Mengchen Liu,[1] Zeyu Hui,[1] Paulina J. Ai,[5] Kent Griffith,[4] Chunsheng Wang,[3]* Shyue Ping Ong,[1]* Yan Yao,[2]* Ping Liu[1]*

**Corresponding author:** Ping Liu, piliu@ucsd.edu; Yan Yao, yyao4@central.uh.edu; Shyue Ping Ong, ongsp@ucsd.edu; Chunsheng Wang, cswang@umd.edu.

[#]These authors contribute equally to this work

[1]Department of Nanoengineering, University of California, San Diego, La Jolla, CA, USA.

[2]Materials Science and Engineering Program and Texas Center for Superconductivity at the University of Houston, University of Houston, Houston, TX, USA.

[3]Department of Chemical and Biomolecular Engineering, University of Maryland, College Park, MD, USA

[4]Department of Chemistry and Biochemistry, University of California San Diego, La Jolla, CA 92093

[5] The Bishop's School, La Jolla, CA, 92037

**Abstract**

Unlike Li-ion transport in the bulk of carbonaceous materials, little is known about Li-ion diffusion on their surface. In this study, we have discovered an ultra-fast Li-ion transport phenomenon on the surface of carbonaceous materials, particularly when they have limited Li insertion capacity along with a high surface area. This is exemplified by a carbon black, Ketjen Black (KB). An ionic conductivity of 18.1 mS cm$^{-1}$ at room temperature is observed, far exceeding most solid-state ion conductors. Theoretical calculations reveal a low diffusion barrier for the surface Li species. The species is also identified as Li*, which features a partial positive charge. As a result, lithiated KB functions effectively as an interlayer between Li and solid-state electrolytes (SSE) to mitigate dendrite growth and cell shorting. This function is found to be electrolyte agnostic, effective for both sulfide and halide SSEs. Further, lithiated KB can act as a high-performance mixed ion/electron conductor that is thermodynamically stable at potentials near Li metal. A graphite anode mixed with KB instead of a solid electrolyte demonstrates full utilization with a capacity retention of ~85% over 300 cycles. The discovery of this surface-mediated ultra-fast Li-ion transport mechanism provides new directions for the design of solid-state ion conductors and solid-state batteries.

**Introduction**

Li-ion transport in carbonaceous materials is an important process, underpinning their applications in rechargeable batteries.[1-4] Up to now, our understanding of this transport process is associated with the Li-ion diffusion in Li-C intercalation compounds.[5,6] The kinetics of such processes are highly dependent on the structures of the carbon materials, which can be categorized into three different types: graphite, hard carbon and soft carbon.[7,8]

Li-ion transport in graphite, which consists of highly ordered, stacked graphene layers, is slow due to the high intercalation/diffusion barriers between the layers.[9,10] In contrast, hard carbons feature short-range graphitized domains, expanded interlayer distance, abundant voids or pores, and rich edges and defects.[11] These characteristics have endowed hard carbons with the ability to uptake Li-ion in nanopores and absorb Li-ion at defect sites, thus enabling fast intercalation/ deintercalation.[12,13] Soft carbons have well-ordered graphene layers but with randomly stacked turbostratic structures,[14] which facilitates lithiation/delithiation.[15,16] Regardless of their structure, these bulk diffusion processes are generally slow. As a result, additional ion conductors, either liquid or solid, are needed to enable these carbon materials to function as electrode materials with reasonable rates.[17,18]

In contrast, the transport of Li on the surface of carbon is virtually unexplored due to two factors. Carbonaceous materials used in batteries tend to have very low surface area, which is necessary to reduce the parasitic reactions and the amount of charge consumed to form the solid electrolyte interface (SEI) in organic electrolytes.[3,19]

In addition, the presence of the SEI also makes the observation of any surface transport processes unfeasible.[20-22] However, understanding the transport of Li on the surface of carbon is both scientifically and technologically important. Scientifically, the chemical state of Li ion on the surface can be significantly different from the bulk. At near the electrochemical potential of Li metal, the Li species on the surface of carbon might be only partially ionized. As a result, its interaction with the carbon surface might be weaker which is essential for fast transport. Technologically, realizing rapid Li transport on carbon surface enables a variety of applications. In addition to being a high-rate electrode material,[23,24] carbon has also been used as an interlayer to protect Li metal electrodes in solid-state batteries by guiding Li deposition away from the SSE interface.[25-27] However, Li-intercalated carbon is usually lithiophilic.[28,29] Consequently, a carbon interlayer needs to have limited Li intercalation capacity but offers fast transport of Li. Finally, there is a pressing need to develop mixed electron/ion conductors that are thermodynamically stable near the potential of Li metal to enable high-rate, long-life anodes for solid-state batteries. Currently, active materials are mixed with carbon as an electronic conducting additive and solid electrolytes to form a three-phase composite structure. However, virtually none of the known solid electrolytes are practically stable during Li plating at a negative potential vs. Li.

Solid-state batteries provide a unique platform to study the Li transport on the surface of carbon because of the absence of an SEI layer. In this regard, we have systematically studied the surface structures of lithiated carbonaceous materials with a variety of structures and surface area values. We have identified a partially charged Li* species on the surface of lithiated carbon black materials. Theoretical calculations reveal a diffusion barrier of Li* on the surface of lithiated carbon as low as 0.149 eV and approximately one-third of the value in the bulk structure. Consequently, an effective ionic conductivity of 18.1 mS cm$^{-1}$ at room temperature is observed, far exceeding most of the solid-state Li-ion conductors. The rapid surface Li* transport mechanism enables lithiated carbon blacks with large surface area values to serve as effective interlayers between Li metal and an SSE layer, greatly improving the cycling stability of Li metal anode in various SSEs systems. Moreover, as a thermodynamically stable mixed conductor, lithiated carbon black is shown to enable a stable graphite anode in the absence of additional solid electrolytes. The discovery of this surface-mediated ultra-fast Li transportation mechanism provides new directions for the materials design of solid ion conductors and solid state batteries.

**Speciation of mobile Li on the surface of carbonaceous materials**

To study the reaction between Li and carbon, we selected six carbon materials with different structures and surface area values (Supplementary Table 1 and Supplementary Fig. 1): graphite (Gr, 2.0 m$^2$ g$^{-1}$), mesocarbon microbeads synthetic graphite (MCMB, 2.2 m$^2$ g$^{-1}$), carbon black (CB, 29.1 m$^2$ g$^{-1}$), Super-P (SP, 57.5 m$^2$ g$^{-1}$), multiwall carbon nanotubes (MWCNTs, 208.3 m$^2$ g$^{-1}$), and Ketjen black (KB, 729.2 m$^2$ g$^{-1}$). Their lithiated forms are labeled as Li-Gr, Li-MCMB, Li-SP, Li-MWCNTs, and Li-KB, respectively. Details for preparing lithiated carbon samples are described in the Methods section. To reveal the chemical environment of Li, Li-Gr, Li-SP and Li-KB are first investigated by solid-state nuclear magnetic resonance spectroscopy (NMR, Fig. 1a). Li-Gr shows a single peak at 43 ppm, which is the typical peak assigned to LiC$_6$.[30] In contrast, Li-SP shows one peak at 22.3 ppm, while Li-KB has two peaks at 6.5 ppm and 1.3 ppm, respectively. To probe the structure and dynamics of Li species, variable temperature NMR experiments are also performed (Supplementary Fig. 2). Compared to the other two samples, the peak centered at 6.5 ppm in Li-KB exhibits much stronger temperature dependence with a reduced FWHM and shorter T$_1$ relaxation time upon the increase of temperatures (Supplementary Fig. 3a-d), indicating increased Li-ion mobility.[31] However, the peak at 1.3 ppm has negligible temperature sensitivity, similar to the peaks for Li-Gr and Li-SP. More interestingly, $^7$Li-$^7$Li 2D-EXSY experiment (Supplementary Fig. 3e) shows the two peaks in Li-KB don't have apparent exchange, indicative of two different Li chemical environments. Therefore, the peak at 1.3 ppm is likely associated with the Li intercalated in the bulk structure of carbon, while the peak at 6.5 ppm is related to a new Li chemical environment, which features a much higher Li-ion mobility upon the increase of temperature.

To further resolve the chemical environment of Li, we turn to Raman and X-ray photoelectron (XPS) spectroscopy. The Raman spectra show that these carbons exhibit very different reactivity towards Li (Fig. 1b). Gr shows a disappearance of its D and G bands upon lithiation, while a new, weak peak close to 900 cm$^{-1}$ arises, which is attributed to the C-C ring breathing mode.[32] Li-SP exhibits a slight intensity decrease in its D and G bands, while the C-C ring breathing peak is also present. Li-MWCNTs displays an even smaller change in its D and G bands with a very weak C-C ring breathing peak. Li-KB, in contrast, doesn't show any change in spectra as compared to pristine KB. The Raman spectra indicate that the bulk reactivity follows this order: Gr > SP > MWCNTs > KB, which is consistent with their intercalation capacities measured in SSBs (Supplementary Fig. 4a). Based on the discharge capacity, the Li:C ratio is calculated to be 1:31 for Li-KB, 1:22 for Li-MWCNTs, 1:15 for Li-SP and 1:6 for Gr, respectively. This bulk reactivity difference is visually evident in Supplementary Fig. 4b, where Li-KB remains black while Li-Gr has a golden color, consistent with the formation of $LiC_6$.[17] The varied bulk reactivity and surface area values of different carbons provide the possibility to study the surface characteristics for Li transportation.

The survey XPS spectra for all samples are shown in supplementary Fig. 4c, where the peaks can be resolved by referring to previous studies.[33,34] Besides the commonly reported $Li_2O_2$, $Li_2CO_3$, and $LiC_6$ (Li-ion hosted by one carbon ring) species, a new peak at 54.8 eV is observed in the spectra of Li-KB, Li-SP, and Li-MWCNTs (Fig. 1c). This peak has a higher binding energy than Li metal (at 53.8 eV), but lower than $LiC_6$, suggesting that this Li* species is less ionized than the Li-ion in $LiC_6$, a state that has not been observed before. Among all the samples, Li-KB has the highest ratio of Li*. Again, this carbon has the highest surface area and lowest intercalation capacity as well. On the contrary, this peak is not observed in Gr. Overall, the crystal structure and surface area of carbon determine the reactivity between Li and carbon as well as the chemical environments of Li on the surface.

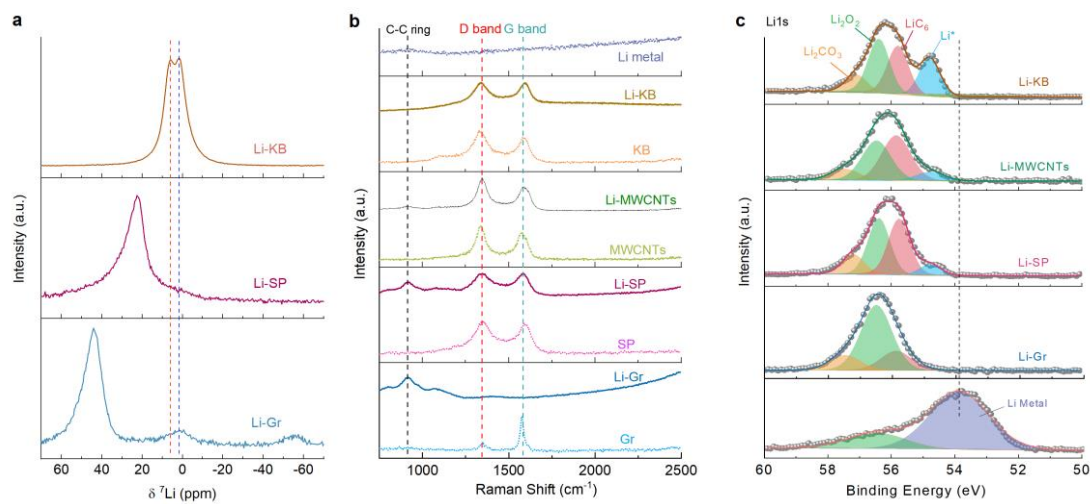

**Fig. 1 | Chemical environment of Li in different carbon materials.** (a) $^7$Li NMR spectra of Li-Gr, Li-SP and Li-KB; (b) Raman spectra of KB, MWCNTs, SP, Gr before and after lithiation; and (c) High resolution Li 1s XPS of Li-KB, Li-MWCNTs, Li-SP and Li-Gr.

Climbing image nudged elastic band (CI-NEB) simulations[35] were conducted to determine the Li* migration barriers on the surface and in the bulk of lithiated carbon. A $2 \times 2 \times 2$ supercell of bulk LiC$_6$ with one Li vacancy and $n_a \times n_b \times 4$ slab supercells of LiC$_6$ with one Li* on surface were used (Fig. 2a), where $n_a$ and $n_b$ are the integer multiples of the a and b lattice parameters. The migration barriers of Li* in the bulk are 0.436-0.453 eV, in line with previous studies[36,37]. In contrast, the Li* diffusion barriers on the surface of Li$_{13}$C$_{96}$ (surface Li/C = 1/24) are 0.222-0.226 eV, nearly half that in the bulk. Such low surface migration barriers of Li* are consistently observed on surfaces with a range of Li/C ratios from 1/54 to 1/12, with a general trend that Li* migration barriers decrease as the surface Li/C ratio increases (Fig. 2c). At a surface Li/C ratio of 1/12, the Li* migration barriers are 0.149-0.160 eV, lower than well-known Li superionic conductors (e.g., $Li_{10}GeP_2S_{12}$ with a diffusion activation barrier of 0.21-0.24 eV)[38,39]. Overall, CI-NEB simulations indicate significantly more facile Li* migration on the surface than in the bulk of lithiated carbon.

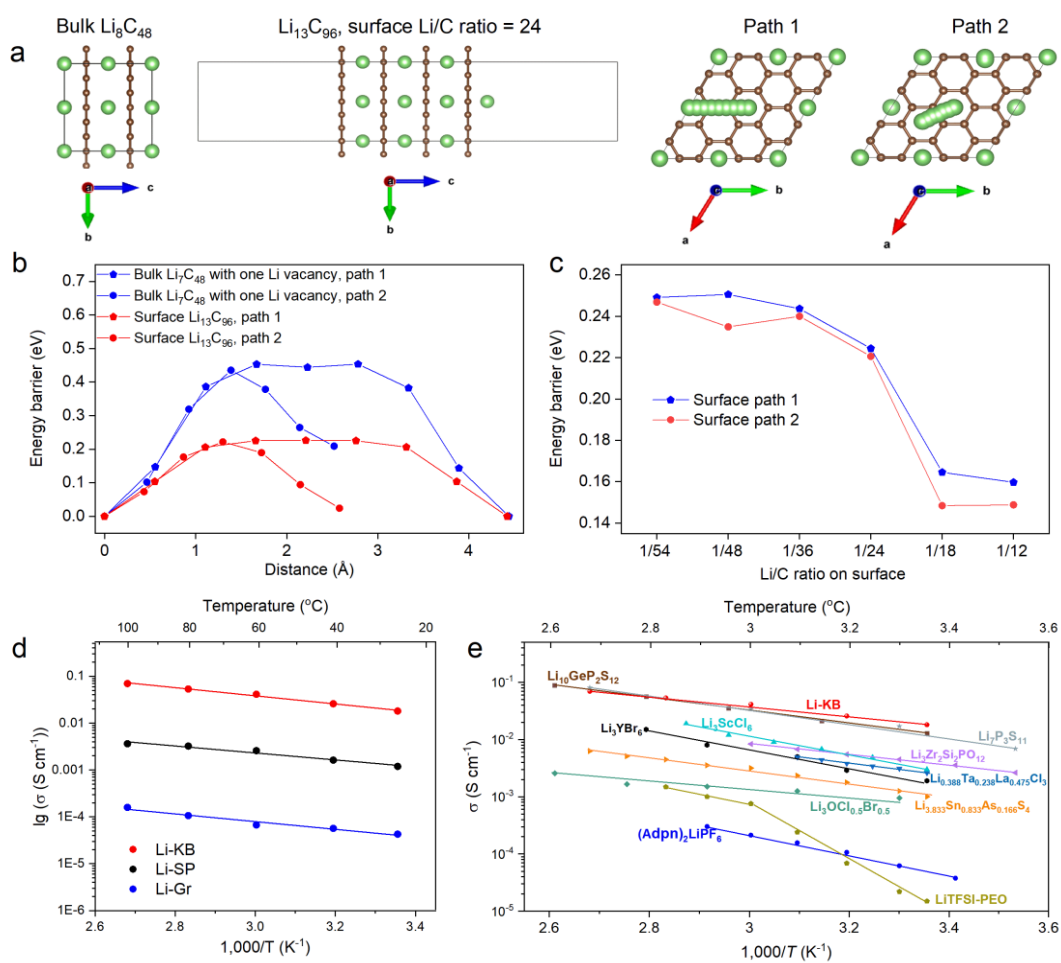

**Fig. 2 | Surface Li* migration barriers and ion conductivity of lithiated carbons.** (a) Bulk and surface structures of lithiated carbon and two migration pathways of Li*. (b) CI-NEB energy landscapes for the two types of Li* migration paths in bulk $Li_7C_{48}$ with one Li vacancy and on the surface of $Li_{13}C_{96}$ with surface Li/C ratio of 1/24. (c) CI-NEB Li* migration energy barriers with respect to Li/C ratios on the surface of lithiated carbon. (d) The ionic conductivities of Li-Gr, Li-SP and Li-KB at different

temperatures. (e) Ionic conductivity comparison between Li-KB and other solid-state ionic conductors.

To directly measure the transport of Li through these carbon materials, we construct a symmetric solid-state battery cell with a structure of Li/C/LPSCl/C/Li. Based on the resistance differences between symmetric cells with different C layer thicknesses (Supplementary Fig. 5-7 and Table 2) as determined by a direct current (DC) method, the ionic resistance of carbon interlayers can be calculated which is in turn used to estimate their ionic conductivities. Our method assumes Li goes through carbon interlayers and is deposited between the carbon layer and Li, which is proven later. We note that the DC method is more appropriate than the commonly used electrochemical impedance method due to the high electronic conductivity of these materials[40]. More detailed discussions and calculation procedures are shown in Methods. As shown Fig. 2d, Li-KB has Li ionic conductivity values of 18.1, 25.8, 41.1, 53.2 and 69.6 mS cm$^{-1}$ at 25, 40, 60, 80 and 100 ºC, respectively. Based on the Arrhenius formula, the activation energy for Li* migration in Li-KB is calculated to be 0.17 eV, which matches well with the activation barriers predicated by CI-NEB with a surface Li/C ratio from 1/18 to 1/12. Moreover, the ionic conductivities of Li-KB are 10x higher than Li-SP and almost $10^3$x higher than Li-Gr at the same temperature, consistent with our hypothesis that surface Li transport plays a dominant role. More importantly, Li-KB exhibits the highest ionic conductivity at room temperature compared to previously reported solid-state Li-ion conductors (Fig. 2e)[38,41-49], demonstrating the power of the surface Li-ion transport mechanism.

**Correlation between surface Li transport and interlayer function to protect Li metal anode in SSBs**

To explore the functions of ultrafast surface Li* transport on carbon, we first use them as interlayers between Li and SSE layers to circumvent the challenges of Li metal anode in SSBs. Li/C/LPSCl/C/Li symmetric cells are constructed to measure the critical current density (CCD), which is defined as the current density when the cell would short. As an example, the CCDs of cells with KB interlayers are measured with stepwise increase in current densities (from 0.1 to 50 mA cm$^{-2}$). As shown in Supplementary Fig. 8, with the increase of temperature from 25 to 100 °C, the CCDs of KB interlayers increase dramatically from 2.4 to 32.0 mA cm$^{-2}$, suggesting the exceptional ability of KB to enable Li transport while mitigating dendrite penetration. The CCDs of the other 5 different carbon interlayers (Supplementary Fig. 9-13) are summarized in Fig. 3a. At the same temperature, there is an apparent positive correlation between surface area and CCD. This observation is consistent with the surface mediated Li transport mechanism as illustrated in Fig. 1 and Fig. 2. High surface area carbon has prominent surface transport mechanism with limited Li insertion into its bulk. For each specific carbon, CCD increases with temperature, which is likely a result of increased ionic conductivity for the interlayers.

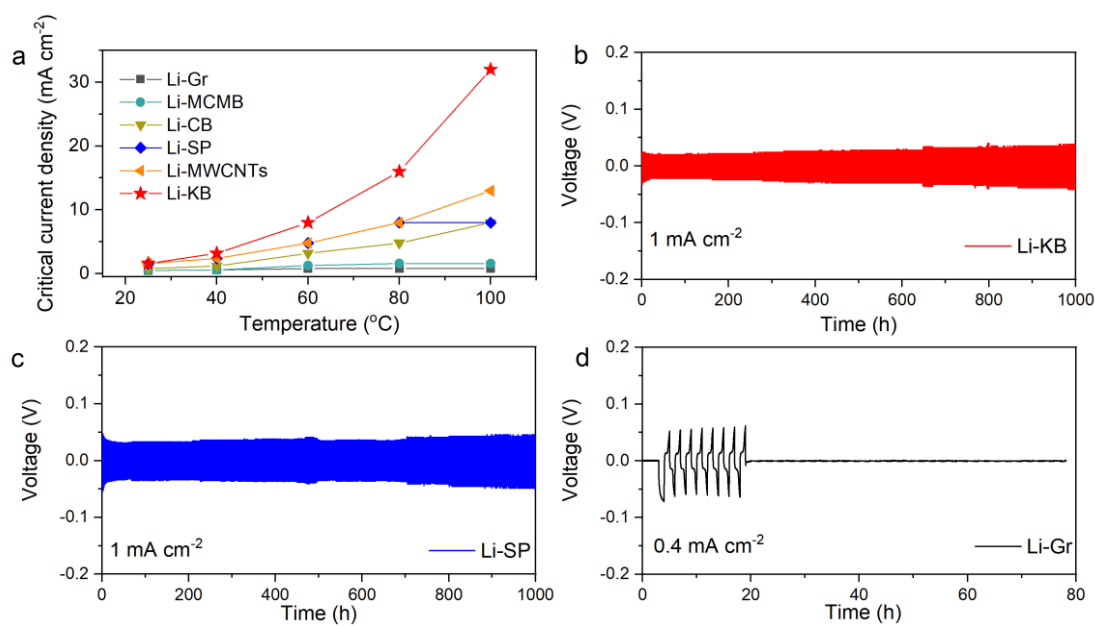

**Fig. 3 | Lithiated carbon as an interlayer to protect Li metal anode in SSBs.** (a) CCDs of carbon interlayers in the Li/C/LPSCl/C/Li symmetric cells. The cycling performance of (b) KB interlayer at 1.0 mA cm$^{-2}$, (c) SP interlayer at 1.0 mA cm$^{-2}$, and (d) Gr interlayer at 0.4 mA cm$^{-2}$.

The cycling stability of Li/C/LPSCl/C/Li symmetric cells are also evaluated at 60 °C. At a current density of 1.0 mA cm$^{-2}$, cells with KB and SP interlayers (Fig. 3b-c) cycle stably over 1000 h without short circuit. In contrast, the Gr interlayer-based symmetric cell is shorted even at 0.4 mA cm$^{-2}$ after cycling for 16 h (Fig. 3d), again indicating the significance of the surface transport mechanism for the carbon layer to protect the Li metal anode. Moreover, the KB interlayer-based symmetric cell cycled at 60 °C exhibits a negligible increase of impedance over 1000 h (Supplementary Fig. 14), indicative of good chemical compatibility between the carbon interlayer and LPSCl.

To visualize the deposition behavior of Li, we captured cross-sectional SEM images of the Li/C/LPSCl/C/Li cells after plating/stripping Li metal (Fig. 4). The images of the cell with a Gr interlayer after undergoing 10 hours of cycling at 0.2 mA cm$^{-2}$ and 40 °C (Supplementary Fig. 15a), are displayed in Fig. 4a-c and Supplementary Fig. 16 and 17. While the contact between Gr and Li is good, we observe a small amount of Li deposited between LPSCl and Gr. This deposition likely serves as a starting point to induce more deposition at this interface, leading to Li dendrite formation and short circuit of the cell. The deposition of Li at the electrolyte/Gr interface is likely due to the sluggish Li transport through the bulk of the graphite structure. SEM images of a symmetric cell with an SP interlayer are shown in Fig. 4d-f and Supplementary Fig. 18-19. The cell underwent 5.0 mAh cm$^{-2}$ of cycling at 1.0 mA cm$^{-2}$ (Supplementary Fig. 15b) and 40 °C. The images reveal that the SP interlayers maintain excellent contact with LPSCl and Li on both sides, which ended their cycling with the final step being stripping and plating, respectively. The SP interlayer appears to have successfully prevented the formation of Li dendrites. All of the Li is plated underneath the SP interlayer. The Li layers are very dense: there is a thickness difference of ~50 μm between the two sides (Supplementary Fig. 18a and 19a), matching well with the expected Li thickness difference (~50 μm for moving 5 mAh cm$^{-2}$ from one side to the other). Similar results are obtained with the use of KB interlayers as shown in Fig. 4g-i and Supplementary Fig.20-21. The symmetric cell is cycled at 40 °C for 5 hours at 1 mA cm$^{-2}$ (Supplementary Fig. 15c). In cells with SP or KB interlayers, Li appears to go through the carbon interlayers without depositing at the electrolyte/carbon interface or inside the interlayer. Instead, Li preferentially deposits at the Li/interlayer interface. Such a phenomenon is attributed to the high ionic conductivity of the carbon. Despite the fact that carbon is electronically conductive, the differences in lithium nucleation on carbon versus on lithium are sufficient to drive preferential lithium deposition at the carbon/lithium interface.

To further illustrate the relationship between CCD and Li deposition patterns, we also characterized the symmetric cells cycled above their CCDs (Supplementary Fig. 22-26). When the cell with the SP interlayers is cycled at 40 °C and 6 mA cm$^{-2}$, there is dendritic Li deposition occurring between the SP interlayer and LPSCl, leading to the failure of the cell (Supplementary Fig. 23). On the stripping side, the LPSCl/SP interface is still well maintained, but the contact at the SP/Li interface is poor (Supplementary Fig. 24). In the case of cells with KB interlayers, Li is also found to deposit between KB and LPSCl (Supplementary Fig. 25-26) as the current density is increased to 10 mA cm$^{-2}$ at 40 °C. These findings show that the preferential deposition at the Li/interlayer interface enabled by the differences in nucleation energy can be compromised at high current densities. When the transport resistance of Li through the interlayer is larger than the differences in nucleation energy, Li will deposit at the electrolyte/interlayer interface.

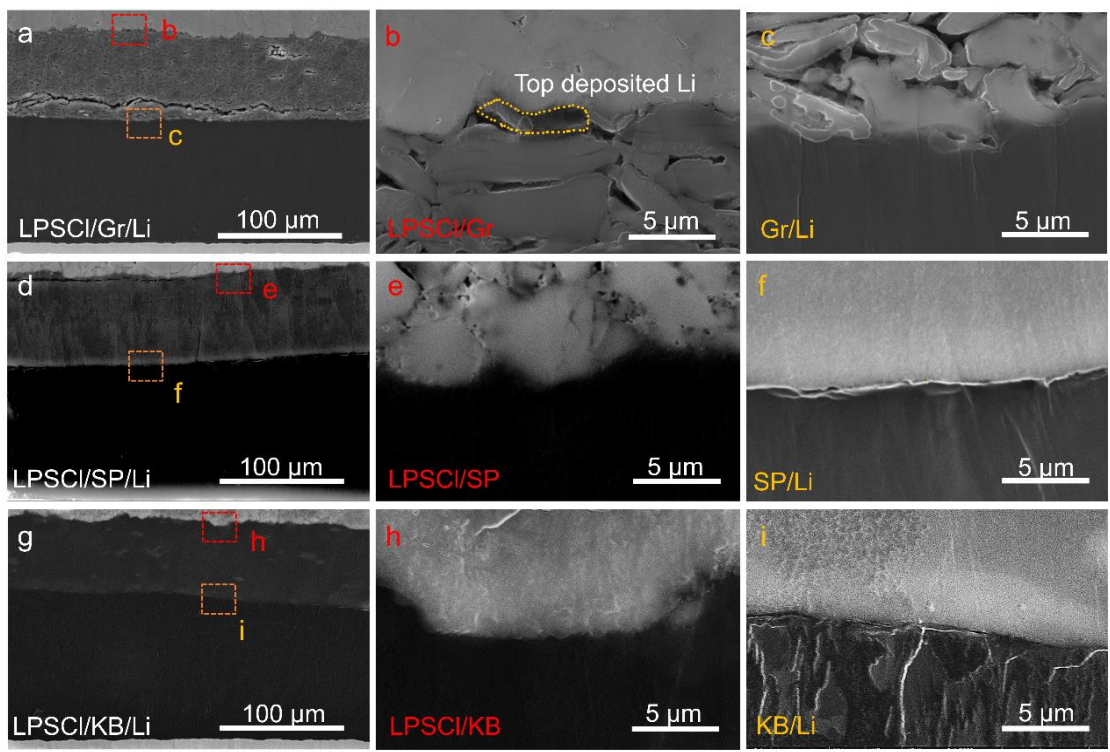

**Fig. 4 | Morphology of Li/C/LPSCl interfaces after cycling.** Cross-sectional SEM images of Li/C/LPSCl interfaces after plating Li metal with different C: (a) Gr, 2.0 mAh cm$^{-2}$, 0.2 mA cm$^{-2}$; (b) SP, 5.0 mAh cm$^{-2}$, 1.0 mA cm$^{-2}$; and (c) KB, 5.0 mAh cm$^{-2}$, 1.0 mA cm$^{-2}$.

**Application of KB as an interlayer in SSBs with different electrolytes**

We have identified KB as a very promising material to provide rapid surface Li transport. Next, we evaluate its suitability as an interlayer material in SSBs to improve Li metal anode stability. A self-standing KB@MWCNTs film is first fabricated. The film features a KB: MWCNTs weight ratio of 7:3 and a mass loading of 1.5 mg cm$^{-2}$ (Supplementary Fig. 27a). The film thickness is ~11 μm (Supplementary Fig. 27b). When evaluated as an interlayer in a symmetric cell, the CCD is 8.0 mA cm$^{-2}$ at 60 °C (Supplementary Fig. 27c), which is the same as that is obtained with a pure KB interlayer. A full cell with a structure of NMC811/LPSCl/KB@MWCNTs/Li is fabricated. After a formation cycle at 10 mA g$^{-1}$, the cell is then cycled at a rate of 200 mA g$^{-1}$. As shown in Fig. 5a, the cell achieves a capacity retention of ~85% after 300 cycles. In contrast, the cell without the interlayer is short-circuited in the first cycle at a current density of 100 mA g$^{-1}$ (Supplementary Fig. 28a). Even at a lower rate of 20 mA g$^{-1}$, the cell without the interlayer is also short-circuited by the 21$^{th}$ cycle (Supplementary Fig. 28b).

In principle, the ability of carbon interlayer to suppress Li dendrite growth is agnostic to the electrolyte material as long as the electrolyte/carbon interface is stable. In this regard, we also fabricated both symmetric and full cells by replacing LPSCl with Li$_3$YCl$_6$ (LYC). As expected, the symmetric cell cycled stably for over 1000 hours at 0.2 mA cm$^{-2}$ (Supplementary Fig. 29). A full cell can be cycled at a rate of 200 mA g$^{-1}$ for 50 cycles without short circuit (Fig. 5c-d). In contrast, the control cell without the interlayer is short-circuited at the 5th cycle even at a much lower rate of 50 mA g$^{-1}$ (Supplementary Fig. 28c). The ability of the interlayer to protect Li by enabling preferential deposition at the interlayer/Li interface is very encouraging, considering that halide electrolytes are well known to be unstable with Li metal.[50]

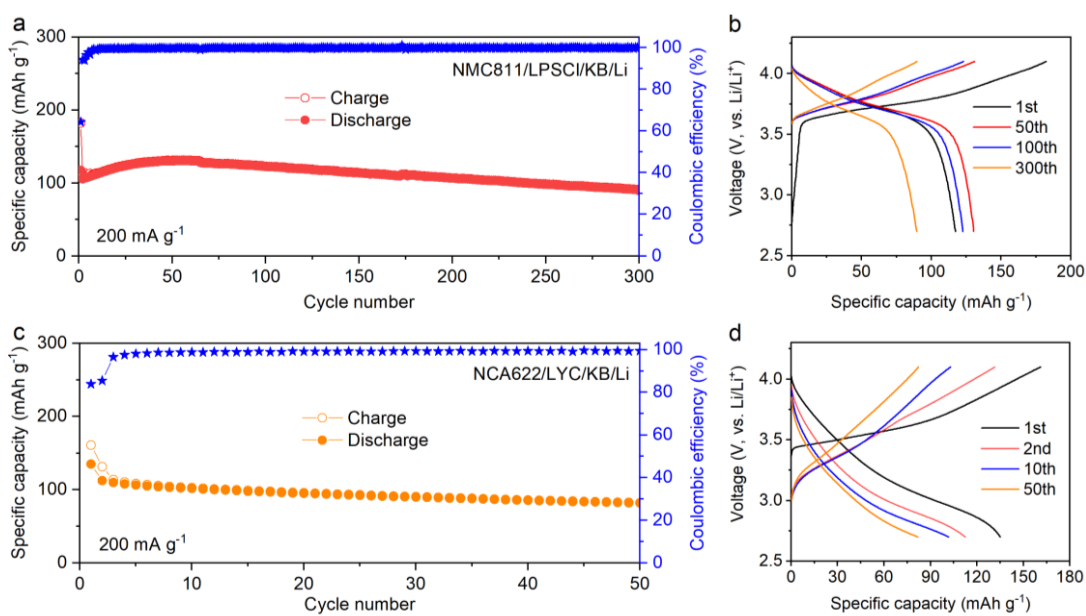

**Fig. 5 | Electrochemical performance of full cells with different electrolytes enabled by the Li-KB interlayer.** (a) NMC811/LPSCl/KB@MWCNTs/Li cell cycling performance at 60 ºC and at 200 mA g$^{-1}$ and (b) the corresponding voltage profiles. (c) NCA622/LYC/KB@MWCNTs/Li cell cycling performance at 60 ºC and at 200 mA g$^{-1}$, and (d) the corresponding voltage profiles.

**Application of KB as a mixed conductor for solid-state electrodes**

KB is further explored as a thermodynamically stable, mixed electronic and ionic conductor to enable stable anodes in SSBs. Here, MCMB is evaluated as an anode material for an SSB cell. Instead of adding a solid electrolyte into the anode composite, we add KB to serve the function of the electrolyte. Previously, sulfide solid electrolyte has been shown to suffer from significant decomposition at the working potentials of MCMB.[17,51] The ability of KB to enable rapid Li transport is first demonstrated in a chemical lithiation experiment. MCMB, with or without added KB, is directly pressed on a piece of Li metal. Time dependent XRD of MCMB is then used to monitor the degree of lithiation. Supplementary Fig. 30 shows that, with the addition of KB, the lithiation of MCMB is much faster, confirming the ability of KB to transport Li-ions in the mixture. As shown in Fig. 6a, MCMB electrode containing 10 wt% of KB displays much higher rate capabilities than pure MCMB or MCMB mixed with LPSCl electrodes in MCMB/LPSCl/KB/Li half-cells. It should be noted that the mass loading of MCMB is ~10 mg cm$^{-2}$. The cells are tested at rates from 37 to 1110 mA g$^{-1}$ at a temperature of 60 $^{o}$C. The voltage profiles of the MCMB@KB anode show the signature voltage plateaus for graphite (Fig. 6b), indicating the much more favorable electrode kinetics compared to the other two electrodes where these features are absent (Supplementary Fig. 31a-b). The long-term cycling stabilities of the three carbon anodes are measured at a rate of 74 mA g$^{-1}$ at 60 $^{o}$C (Fig. 6c). MCMB@KB anode shows a capacity retention of ~85% after 300 cycles with an average coulombic efficiency of over 99.6%. In contrast, the MCMB and MCMB@LPSCl electrodes suffer from shorting at the 87th and 212th cycle, respectively. Further, MCMB@LPSCl shows much lower coulombic efficiency than MCMB@KB, especially during the initial 50 cycle (Fig. 6d and Supplementary Fig. 31c), which is attributed to the poor stability of LPSCl at the working potential of MCMB anode. Therefore, KB is a thermodynamically stable, high-conductivity material, well-suited for high-performance SSBs anodes.

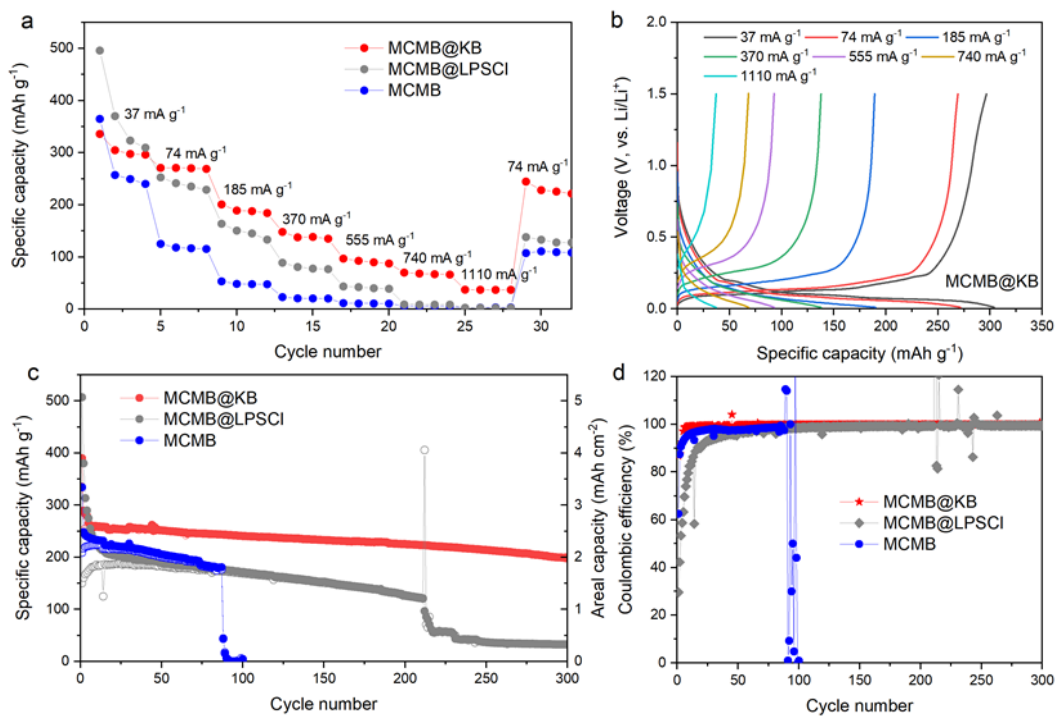

**Fig. 6 | Using Lithiated KB as a mixed conductor in solid-state MCMB anodes.** (a) Rate performance of pure MCMB, MCMB@KB and MCMB@LPSCl at current densities from 37 to 1110 mA g$^{-1}$ and at 60 °C. (b) Corresponding voltage profiles of MCMB@KB anode at different rates. (c) The cycling performance of pure MCMB, MCMB@KB and MCMB@LPSCl anodes at 74 mA g$^{-1}$ and at 60 °C in SSBs. (d) The corresponding coulombic efficiencies in (c).

In summary, we have discovered a new surface mediated Li-ion transport mechanism in carbon materials. For carbon blacks with high surface area and low intercalation capacity, lithiation results in materials with high ionic conductivities and low migration barriers. These carbon materials have shown great potential in SSBs either as interlayers to protect the Li metal anode or as a thermodynamically stable mixed conductor for constructing other anodes (e.g., MCMB). Their ability to serve as an interlayer to enable preferential deposition at the interlayer/Li interface also appears to be electrolyte agnostic. We have shown their effectiveness for both sulfide and halide-based electrolytes. The discovery of this surface mediated ultra-fast Li-ion transportation mechanism provides a new direction for the design of solid-state ionic conductors with high conductivities and the practical applications of SSBs.

**Acknowledgments** This work is supported by the Advanced Research Projects Agency-Energy, U.S. Department of Energy (DOE), under Contract No. DE-AR0000781.

**Author contributions**
J.Z. and P.L. conceptualized the idea and designed all of the experiments. S.W. did the XPS and Rama tests and results analysis. Chaoshan Wu and Y.Y. collected the cross-sectional SEM and EDX mapping images. J.Q., T.W.K. and S.P.O. performed the theoretical calculations. H.W. and Chunsheng Wang tested the full cell performance. S.L. and K.G. performed the NMR measurement. N.H., S.F. Z.L. H.L., N.S., C.S. Z.H. and P.A. helped with the materials preparation, cell assembly, results discussion and manuscript revision. J.Z. and P.L. drafted the manuscript with the input and revision from all authors. Chunsheng Wang, S.P.O, Y.Y., and P.L. supervised this research. J.Z., S.W., Chaoshan Wu, J.Q. and H.W. contribute equally to this work.

**Competing interests**
The authors declare no competing interests.

**References:**
1    Fong, R., von Sacken, U. & Dahn, J. R. Studies of lithium intercalation into carbons using

| | |
|---|---|
| | nonaqueous electrochemical cells. *J. Electrochem. Soc.* **137**, 2009-2013 (1990). |
| 2 | Manthiram, A. An Outlook on Lithium Ion Battery Technology. *ACS Cent Sci* **3**, 1063-1069, doi:10.1021/acscentsci.7b00288 (2017). |
| 3 | Xu, K. Electrolytes and interphases in Li-ion batteries and beyond. *Chem Rev* **114**, 11503-11618, doi:10.1021/cr500003w (2014). |
| 4 | Roberts, A. D., Li, X. & Zhang, H. Porous carbon spheres and monoliths: morphology control, pore size tuning and their applications as Li-ion battery anode materials. *Chem Soc Rev* **43**, 4341-4356, doi:10.1039/c4cs00071d (2014). |
| 5 | Xie, L. *et al.* Hard Carbon Anodes for Next‐Generation Li‐Ion Batteries: Review and Perspective. *Advanced Energy Materials* **11**, doi:10.1002/aenm.202101650 (2021). |
| 6 | Kaskhedikar, N. A. & Maier, J. Lithium Storage in Carbon Nanostructures. *Adv Mater* **21**, 2664-2680, doi:10.1002/adma.200901079 (2009). |
| 7 | Adams, R. A., Varma, A. & Pol, V. G. Carbon Anodes for Nonaqueous Alkali Metal‐Ion Batteries and Their Thermal Safety Aspects. *Advanced Energy Materials* **9**, doi:10.1002/aenm.201900550 (2019). |
| 8 | W.J. Cao, J. S. Z., D. Adams and J.P. Zheng. Comparative Study of the Power Performance for Advanced Li-Ion Capacitors with Various Carbon Anodes. *ECS Transactions* **61**, 37-48. (2014). |
| 9 | Ping Yu, B. N. P., J. A. Ritter, and R. E. White. Determination of the Lithium Ion Diffusion Coefficient in Graphite. *Journal of The Electrochemical Society* **146**, 8-14 (1999). |
| 10 | Xie, J. & Lu, Y. C. A retrospective on lithium-ion batteries. *Nat Commun* **11**, 2499, doi:10.1038/s41467-020-16259-9 (2020). |
| 11 | Saurel, D. *et al.* A SAXS outlook on disordered carbonaceous materials for electrochemical energy storage. *Energy Storage Materials* **21**, 162-173, doi:10.1016/j.ensm.2019.05.007 (2019). |
| 12 | Li, Q. *et al.* Unraveling the Key Atomic Interactions in Determining the Varying Li/Na/K Storage Mechanism of Hard Carbon Anodes. *Advanced Energy Materials* **12**, doi:10.1002/aenm.202201734 (2022). |
| 13 | Dahn, E. B. a. J. R. Li-insertion in hard carbon anode materials for Li-ion batteries. *Electrochimica Acta* **45**, 121-130 (1999). |
| 14 | Yu, K. *et al.* Molecular tuning of sulfur doped quinoline oligomer derived soft carbon for superior potassium storage. *Carbon* **191**, 10-18, doi:10.1016/j.carbon.2022.01.034 (2022). |
| 15 | Jian, Z. *et al.* Insights on the Mechanism of Na-Ion Storage in Soft Carbon Anode. *Chemistry of Materials* **29**, 2314-2320, doi:10.1021/acs.chemmater.6b05474 (2017). |
| 16 | Chen, Z., Wang, Q. & Amine, K. Improving the performance of soft carbon for lithium-ion batteries. *Electrochimica Acta* **51**, 3890-3894, doi:10.1016/j.electacta.2005.11.004 (2006). |
| 17 | Xing, X. *et al.* Graphite-Based Lithium-Free 3D Hybrid Anodes for High Energy Density All-Solid-State Batteries. *ACS Energy Letters* **6**, 1831-1838, doi:10.1021/acsenergylett.1c00627 (2021). |
| 18 | Zhou, L., Minafra, N., Zeier, W. G. & Nazar, L. F. Innovative Approaches to Li-Argyrodite Solid Electrolytes for All-Solid-State Lithium Batteries. *Acc Chem Res* **54**, 2717-2728, doi:10.1021/acs.accounts.0c00874 (2021). |
| 19 | Ji, L. *et al.* Controlling SEI formation on SnSb-porous carbon nanofibers for improved Na ion storage. *Adv Mater* **26**, 2901-2908, doi:10.1002/adma.201304962 (2014). |
| 20 | Chen, Y. *et al.* Li metal deposition and stripping in a solid-state battery via Coble creep. *Nature* **578**, 251-255, doi:10.1038/s41586-020-1972-y (2020). |
| 21 | Zhou, S. *et al.* Efficient diffusion of superdense lithium via atomic channels for dendrite-free |
| | lithium–metal batteries. *Energy & Environmental Science* **15**, 196-205, doi:10.1039/d1ee02205a (2022). |
| 22 | Yang, G. *et al.* Iron carbide allured lithium metal storage in carbon nanotube cavities. *Energy Storage Materials* **36**, 459-465, doi:10.1016/j.ensm.2021.01.022 (2021). |
| 23 | Soto, F. A. *et al.* Tuning the Solid Electrolyte Interphase for Selective Li- and Na-Ion Storage in Hard Carbon. *Adv Mater* **29**, doi:10.1002/adma.201606860 (2017). |
| 24 | Yan, W. *et al.* Hard-carbon-stabilized Li–Si anodes for high-performance all-solid-state Li-ion batteries. *Nature Energy* **8**, 800-813, doi:10.1038/s41560-023-01279-8 (2023). |
| 25 | Lee, Y.-G. *et al.* High-energy long-cycling all-solid-state lithium metal batteries enabled by silver–carbon composite anodes. *Nature Energy* **5**, 299-308, doi:10.1038/s41560-020-0575-z (2020). |
| 26 | Ye, L. & Li, X. A dynamic stability design strategy for lithium metal solid state batteries. *Nature* **593**, 218-222, doi:10.1038/s41586-021-03486-3 (2021). |
| 27 | Spencer-Jolly, D. *et al.* Structural changes in the silver-carbon composite anode interlayer of solid-state batteries. *Joule* **7**, 503-514, doi:10.1016/j.joule.2023.02.001 (2023). |
| 28 | Du, P. *et al.* The lithiophobic-to-lithiophilic transition on the graphite towards ultrafast-charging and long-cycling lithium-ion batteries. *Energy Storage Materials* **50**, 648-657, doi:10.1016/j.ensm.2022.05.056 (2022). |
| 29 | Duan, J. *et al.* Is graphite lithiophobic or lithiophilic? *Natl Sci Rev* **7**, 1208-1217, doi:10.1093/nsr/nwz222 (2020). |
| 30 | Maxwell, D. C., O'Keefe, Christopher A., Xu, Chao, Grey, Clare P. C13 NMR study of the electronic structure of lithiated graphite. *Physical Review Materials* **7**, doi:10.1103/PhysRevMaterials.7.065402 (2023). |
| 31 | Kuhn, A. *et al.* Li self-diffusion in garnet-type $Li_7La_3Zr_2O_{12}$ as probed directly by diffusion-induced $Li_7$ spin-lattice relaxation NMR spectroscopy. *Physical Review B* **83**, doi:10.1103/PhysRevB.83.094302 (2011). |
| 32 | Neale, A. R., Milan, D. C., Braga, F., Sazanovich, I. V. & Hardwick, L. J. Lithium Insertion into Graphitic Carbon Observed via Operando Kerr-Gated Raman Spectroscopy Enables High State of Charge Diagnostics. *ACS Energy Lett* **7**, 2611-2618, doi:10.1021/acsenergylett.2c01120 (2022). |
| 33 | Wood, K. N. & Teeter, G. XPS on Li-Battery-Related Compounds: Analysis of Inorganic SEI Phases and a Methodology for Charge Correction. *ACS Applied Energy Materials* **1**, 4493-4504, doi:10.1021/acsaem.8b00406 (2018). |
| 34 | Kiyoshi Kanamura, S. S., Hideharu Takezawa, and Zen-ichiro Takehara. XPS Analysis of the Surface of a Carbon Electrode Intercalated by Lithium Ions. *Chem. Mater.* **9**, 1797-1804 (1997). |
| 35 | Henkelman, G., Uberuaga, B. P. & Jónsson, H. A climbing image nudged elastic band method for finding saddle points and minimum energy paths. *The Journal of Chemical Physics* **113**, 9901-9904, doi:10.1063/1.1329672 (2000). |
| 36 | Liu, Q. *et al.* Kinetically Determined Phase Transition from Stage II (LiC(12)) to Stage I (LiC(6)) in a Graphite Anode for Li-Ion Batteries. *J Phys Chem Lett* **9**, 5567-5573, doi:10.1021/acs.jpclett.8b02750 (2018). |
| 37 | Wang, Z., Ratvik, A. P., Grande, T. & Selbach, S. M. Diffusion of alkali metals in the first stage graphite intercalation compounds by vdW-DFT calculations. *RSC Advances* **5**, 15985-15992, doi:10.1039/c4ra15529g (2015). |

38   Kamaya, N. *et al.* A lithium superionic conductor. *Nat Mater* **10**, 682-686, doi:10.1038/nmat3066 (2011).
39   Mo, Y., Ong, S. P. & Ceder, G. First Principles Study of the Li10GeP2S12 Lithium Super Ionic Conductor Material. *Chemistry of Materials* **24**, 15-17, doi:10.1021/cm203303y (2011).
40   Huggins, R. A. Simple method to determine electronic and ionic components of the conductivity in mixed conductors a review. *Ionics* **8**, 300-313. (2002).
41   Asano, T. *et al.* Solid Halide Electrolytes with High Lithium-Ion Conductivity for Application in 4 V Class Bulk-Type All-Solid-State Batteries. *Adv Mater* **30**, e1803075, doi:10.1002/adma.201803075 (2018).
42   Liang, J. *et al.* Site-Occupation-Tuned Superionic Li(x)ScCl(3+x)Halide Solid Electrolytes for All-Solid-State Batteries. *J Am Chem Soc* **142**, 7012-7022, doi:10.1021/jacs.0c00134 (2020).
43   Yin, Y. C. *et al.* A LaCl(3)-based lithium superionic conductor compatible with lithium metal. *Nature* **616**, 77-83, doi:10.1038/s41586-023-05899-8 (2023).
44   Zhao, Y. & Daemen, L. L. Superionic conductivity in lithium-rich anti-perovskites. *J Am Chem Soc* **134**, 15042-15047, doi:10.1021/ja305709z (2012).
45   Lei Zhu, Y. W., Junchao Chen, Wenlei Li, Tiantian Wang, Jie Wu, Songyi Han, Yuanhua Xia, Yongmin Wu, Mengqiang Wu, Fangwei Wang, Yi Zheng, Luming Peng, Jianjun Liu, Liquan Chen, Weiping Tang. Enhancing ionic conductivity in solid electrolyte by relocating diffusion ions to under-coordination sites. *SCIENCE ADVANCES* **8** (2022).
46   Sahu, G. *et al.* Air-stable, high-conduction solid electrolytes of arsenic-substituted Li4SnS4. *Energy Environ. Sci.* **7**, 1053-1058, doi:10.1039/c3ee43357a (2014).
47   Seino, Y., Ota, T., Takada, K., Hayashi, A. & Tatsumisago, M. A sulphide lithium super ion conductor is superior to liquid ion conductors for use in rechargeable batteries. *Energy Environ. Sci.* **7**, 627-631, doi:10.1039/c3ee41655k (2014).
48   Zhao, Y., Tao, R. & Fujinami, T. Enhancement of ionic conductivity of PEO-LiTFSI electrolyte upon incorporation of plasticizing lithium borate. *Electrochimica Acta* **51**, 6451-6455, doi:10.1016/j.electacta.2006.04.030 (2006).
49   Prakash, P. *et al.* A soft co-crystalline solid electrolyte for lithium-ion batteries. *Nat Mater* **22**, 627-635, doi:10.1038/s41563-023-01508-1 (2023).
50   Li, X. *et al.* Progress and perspectives on halide lithium conductors for all-solid-state lithium batteries. *Energy & Environmental Science* **13**, 1429-1461, doi:10.1039/c9ee03828k (2020).
51   Zhu, Y., He, X. & Mo, Y. Origin of Outstanding Stability in the Lithium Solid Electrolyte Materials: Insights from Thermodynamic Analyses Based on First-Principles Calculations. *ACS Appl Mater Interfaces* **7**, 23685-23693, doi:10.1021/acsami.5b07517 (2015).